\documentclass[a4paper, review]{jds}      

\jdsmonth{January}              
\jdsyear{2022}                  
\jdsvolume{xx}                  
\jdsissue{xx}                   
\jdsdoi{xx.xxxx/xxxxxxxxx}      
\shortauthors{Nguyen, Q. et al.}

\usepackage{amsfonts,amsmath,amssymb,amsthm}
\usepackage{booktabs}
\usepackage{xurl}
\usepackage{lipsum}
\nolinenumbers

\title[An Examination of Olympic Sport Climbing]{\singlespacing \Large An Examination of Olympic Sport Climbing Competition Format and Scoring System}

\author[1]{Quang Nguyen}
\author[2]{Hannah Butler}
\author[1]{Gregory J. Matthews\footnote{Corresponding author. Email: \href{mailto:gmatthews1@luc.edu}{gmatthews1@luc.edu}}}
\affil[1]{Department of Mathematics and Statistics, Loyola University Chicago, Chicago, IL, USA}
\affil[2]{Department of Statistics, Colorado State University, Fort Collins, CO, USA}

\begin{document}
\maketitle
\begin{abstract}
\onehalfspacing
Sport climbing, which made its Olympic debut at the 2020 Summer Games, generally consists of three separate disciplines: speed climbing, bouldering, and lead climbing. However, the International Olympic Committee (IOC) only allowed one set of medals each for men and women in sport climbing. As a result, the governing body of sport climbing, rather than choosing only one of the three disciplines to include in the Olympics, decided to create a competition combining all three disciplines. In order to determine a winner, a combined scoring system was created using the product of the ranks across the three disciplines to determine an overall score for each climber. In this work, the rank-product scoring system of sport climbing is evaluated through simulation to investigate its general features, specifically, the advancement probabilities and scores for climbers given certain placements. Additionally, analyses of historical climbing contest results are presented and real examples of violations of the independence of irrelevant alternatives are illustrated. Finally, this work finds evidence that the current competition format is putting speed climbers at a disadvantage.\@
\end{abstract}


\newpage

\section{Introduction}
\label{sec:sec1}

\subsection{Combined Competition Format}
\label{sec1pt1}

The 2020 Summer Olympics in Tokyo, Japan marked the first appearance of
sport climbing on the Olympic stage \citep{ioc2016}. This sport is broken down into
three distinct disciplines: speed climbing, bouldering, and lead
climbing. However, rather than granting a separate sets of medals for
each discipline, the International Olympic Committee (IOC) only allowed
one set of medals each for men and women. As a result, rather than
choosing only one of the three concentrations, all three disciplines were 
included together forming one single combined event. Under this
``triathlon'' format, every climber must compete in all three
concentrations, and their individual score is determined as the product
of the ranks across the three disciplines, with the lowest rank product
declared the winner.

The first discipline, speed climbing, takes place on a standardized 15
meter-high wall where the racers get one chance to try to reach the top
of the wall as quickly as possible. At Tokyo 2020, speed climbing is
being contested in a head-to-head format and under a single elimination
bracket tournament structure. Next, in bouldering, contestants have a
fixed amount of time to attempt to reach the top of a climbing problem
on a 4.5 meter-high wall in as few attempts as possible, without the use
of ropes. Ties are further broken by the number of ``zone holds,'' which
are holds approximately half way through each course. Finally, in lead
climbing, each athlete is given six minutes and one attempt to climb as
high as they can on a wall of height 15 meters. A climber gets one point
for each hold they reach, and the participant that manages to reach
the highest point on the wall wins the lead discipline. If there is a
tie, the competitor with the fastest elapsed time wins.

The decision to combine the three climbing events and only award one set
of medals each for men and women in the Olympics has received a large
amount of criticism from climbing athletes all over the world. In a
series of interviews conducted by Climbing Magazine in 2016 \citep{blanchard2016}, 
a number of climbers shared their thoughts and concerns about the
new Olympic climbing format. Legendary climber Lynn Hill compared the
idea of combining speed climbing, bouldering, and lead climbing to
``asking a middle distance runner to compete in the sprint.'' She then
added, ``Speed climbing is a sport within our sport.'' Other climbers
also hold the same opinion as Hill regarding speed climbing, using words
and phrases like ``bogus,'' ``a bummer,'' ``less than ideal,'' ``not in
support,'' and ``cheesy and unfair'' to describe the new combined
format. Courtney Woods stated, ``Speed climbers will have
the biggest disadvantage because their realm isn't based on difficult
movements.'' Mike Doyle believed, ``Honestly, the people that will
suffer the most are the ones that focus only on speed climbing. Those
skills/abilities don't transfer as well to the other disciplines.'' The
climbers also expressed their hope for a change in the competition
format in future climbing tournaments.

\subsection{Rank-product Scoring}
\label{rank-product-scoring}

At the 2020 Summer Olympics, both sport climbing competitions for male and female begin with 20 climbers who have previously qualified for the Olympics from qualifying events held in 2019 and 2020. All 20 athletes compete in each of the three disciplines in the qualification round, and their performances in each concentration are ranked from 1 to 20. A competitor's combined score is computed as the product of their ranks in each of the three events; specifically,
\begin{equation} \label{eq:1}
Score_i = R^S_i\times R^B_i\times R^L_i,
\end{equation} 
where $R^S_i$, $R^B_i$, and $R^L_i$ are the ranks of the $i$-th competitor in speed climbing, bouldering, and lead climbing, respectively.  

The 8 qualifiers with the lowest score in terms of product of ranks across the three disciplines advance to the final round, where they once again compete in all three events. Similar to qualification, the overall score for each contestant in the final stage is determined by multiplying the placements of speed, bouldering, and lead disciplines, and the athletes are ranked from 1 to 8. The climbers with the lowest, second lowest, and third lowest product of ranks in the final round win the gold, silver, and bronze medals, respectively. This type of rank aggregation method heavily rewards high finishes and relatively ignores poor finishing results. For instance, if climber A finished 1st, 20th, and 20th and climber B finished 10th, 10th, and 10th, climber B would have a score of 1000 whereas climber A would have a much better score of 400, despite finishing last in 2 out of 3 of the events.  

To the best of our knowledge, we know of no sporting event, team or individual, that uses the product of ranks to determine an overall ranking. There are examples of team sports that use the rank-sum scoring to determine the winning team such as cross country where the squad with the lowest sum of ranks of the top five runners is awarded with a first place finish. \cite{hammond2007}, \cite{mixon2012}, \cite{Boudreau2014}, \cite{boudreau2018}, and \cite{Medcalfe2020} pointed out several problems with rank-sum scoring in cross country, most notably, violations of social choice principles. In addition, some individual sports such as the decathlon and heptathlon rely on a sum of scores from the ten or seven events. However, these scores are not determined based on the ranks of the competitors. That is, a decathlete's score is entirely based upon their own times, distances, and heights, and their overall score will remain the same regardless of the performance of other individuals \citep{westera2006}. On the other hand, there are other individual competitions consisting of several events combined, such as crossfit, that do base their scoring on ranks. In each event, points are earned based on the competitor's rank in the event based on a scoring table, and a contestant's final score is based on the sum of their scores across all the events \citep{crossfit2021}. 

To date, there are several articles that have proposed alternative rank aggregation methods to the sport climbing combined format. First, \cite{Parker2018} investigated a variety of approaches to rank and score climbing athletes, namely, geometric mean, Borda count, linear programming, geometric median, top score method, ABS10, and what they refer to as the "merged method." Eventually, they recommended the merged method, which is a combination of the top score and ABS10 methods as a reasonable scoring approach for climbing due to predictive power and satisfaction of social choice theory. More recently, \cite{stin2022} compared the current rank-product scoring method with rank-sum scoring and observed a dramatic change to the outcome of the men's sport climbing final at the 2020 Olympics when the sum of discipline rankings was used instead of the product. Specifically, Tomoa Narasaki, who finished fourth overall at Tokyo 2020, would have won the gold medal if rank-sum was used as the method of scoring instead of rank-product, Furthermore, they proposed an alternative ranking-based scoring scheme that computes the sum of the square roots of each climber's rankings as their overall score.

As a side note, there are applications of the rank-product statistic in other fields. For example, the rank-product statistic can be used to analyze replicated microarray experiments and identify differentially expressed genes. See \cite{Breitling2004} for more information on this statistical procedure.

In this paper, we perform statistical analysis to investigate the limitations of sport climbing's combined competition format and ranking system. We will evaluate whether the concerns of the professional climbers were valid. The manuscript is outlined as follows. We first begin with a simulation study to examine the key properties of rank-product scoring in sport climbing in Section \ref{sec:sec2}. Our analyses of past climbing tournament data are then presented in Sections \ref{sec:sec3}. Finally, in Section \ref{sec:sec4}, we provide a summary of our main findings as well as some discussion to close out the paper.

\section{Simulation Study}
\label{sec:sec2}

In this section, we perform a simulation study to examine the rankings
and scoring for climbers in both qualification and final rounds. There
are two crucial assumptions to our simulation approach:

\vspace{1mm}

\begin{itemize}

\item
  \textbf{Uniform ranks}: The ranks within each discipline follow a
  discrete uniform distribution with lower and upper bounds of
  \([1, 20]\) and \([1, 8]\) for the qualification and final rounds,
  respectively.
  
\vspace{1mm}

\item
  \textbf{Correlation between events}: We want to introduce dependence
  in the ranks between the disciplines. In particular, based on the
  claim that ``Speed climbing is a sport within our sport'' mentioned in Section \ref{sec:sec1}, a correlation between bouldering and lead climbing is
  assumed, whereas speed climbing is assumed completely independent of
  the other two events.
  
\end{itemize}

\vspace{1mm}



In order to generate data that satisfy the assumptions listed above, we performed simulations using the method of copulas.   By definition, a copula is a multivariate distribution function with standard uniform univariate margins \citep{copbook}. In this simulation study, bouldering and lead ranks were considered to have discrete uniform marginal distributions with a non-zero, positive correlation between the ranks. We chose Kendall's $\tau$ \citep{kendall1938} as our measurement of association between the bouldering and lead disciplines. Our examination considers five different levels of correlation between bouldering and lead climbing: \(0\) (no correlation), \(0.25\) (low), \(0.5\) (moderate), \(0.75\) (high), and \(1\) (perfect). 

Using functions from the \texttt{R} package \texttt{copula} \citep{coppkg}, a copula with the desired correlation structure was first calibrated for each correlation value. A random sample of either 20 (for qualification) or 8 (for final) values from the copula was then drawn and the values were ranked within each discipline. Finally, this process was repeated \(10000\) times for each correlation
level. After the simulations were complete, the total scores for every
simulated round and the final standings for the climbing athletes were
calculated. The simulation results allow us to explore different
properties of the sport climbing rank-product scoring system, including
the distributions of total score for qualifying and final rounds, and
the probabilities of advancing to the final stage and winning a medal, given
certain conditions.

\begin{figure}
\centering
\includegraphics{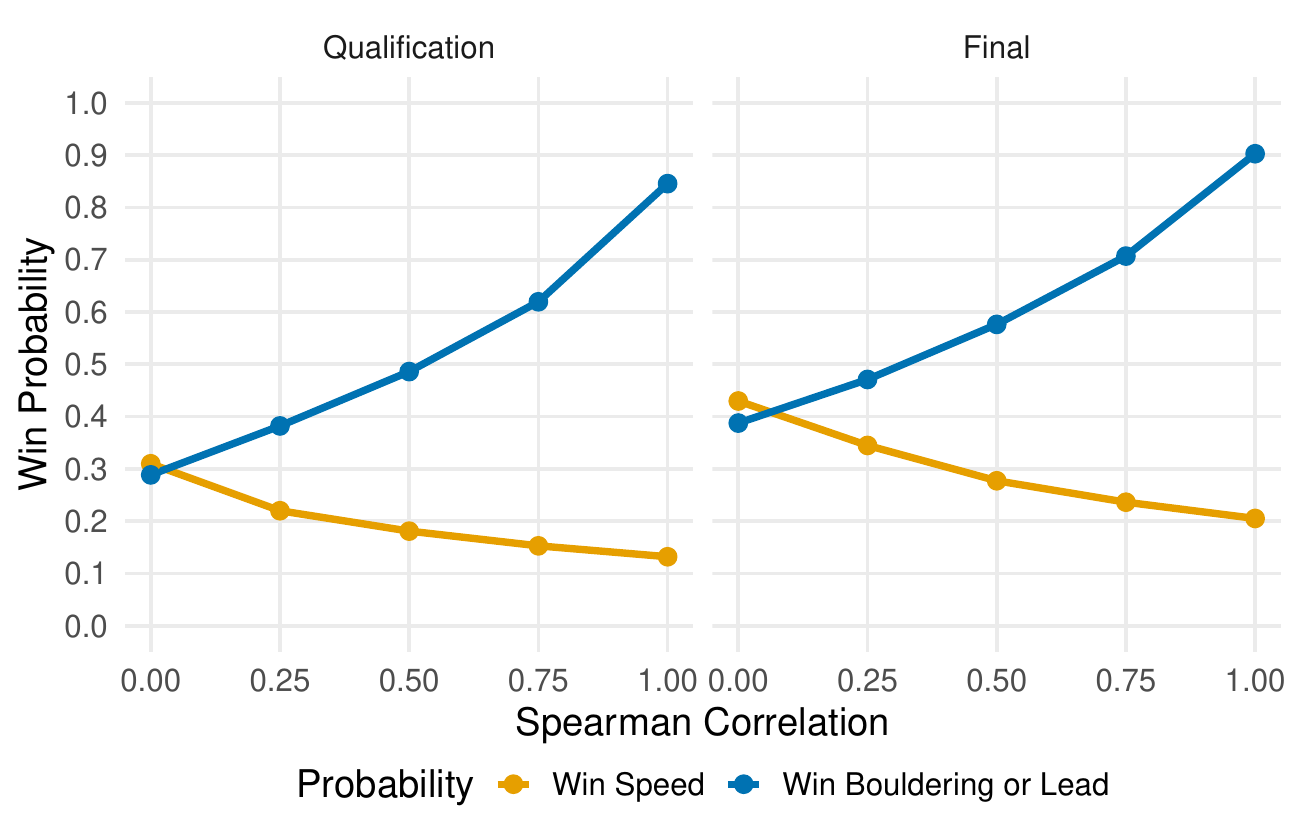}
\caption{\label{fig:fig1}The probabilities of winning qualification
(i.e.~finishing first in qualification) and final rounds (i.e.~winning a
gold medal), given that a climber finishes first in speed versus
finishes first in bouldering or lead. Results were obtained from
simulations for each Kendall correlation value.}
\end{figure}

After obtaining the simulation output, we summarized the advancement
probabilities for both qualifying and final rounds, looking for trends
over the correlation values. Figure \ref{fig:fig1} displays the
probability of finishing first in the qualification and final rounds for each of
the 5 correlation values we considered, color coded by whether a climber
wins the speed event or wins one of bouldering and lead concentrations.
It is easy to see that for both rounds, as the correlation magnitude
increases, the win probability given a first-place finish for lead
climbing or bouldering also goes up. In contrast, the probability of
winning each round given a speed victory decreases as the correlation
between bouldering and lead climbing increases. Moreover, if the two
disciplines lead and bouldering are perfectly correlated, the top-ranked
speed contestant in the final stage only has a 20.5\% chance to win gold,
compared to a 90.3\% chance for a finalist that ranks first in
bouldering and lead climbing. This suggests a potentially unfair
disadvantage for speed specialists when participating in a competition
structure that, as many believed, possesses little skill crossover
between speed and the other two climbing concentrations.


To further examine the advancement probabilities and scores for Olympics
climbing, we implemented the same copula simulation procedure based on empirical results from past climbing competition data. In particular, we used the women's climbing event at Tokyo 2020 as a case study, and obtained Kendall rank correlation coefficients between bouldering and lead climbing of $0.526$ and $0.214$ for the women's qualification and final rounds, respectively. We are specifically interested in the following questions:

\vspace{1mm}

\begin{itemize}
\item
  For a qualifier, what is the probability that they advance to the
  final round (i.e.~finish in the top 8 of the qualification round), given that
  they win any discipline?

\vspace{1mm}  
  
\item
  For a finalist, what is the probability that they win a medal
  (i.e.~finish in the top 3 of the final round), given that they win any
  discipline?
\end{itemize}

\vspace{1mm}

\begin{figure}
\centering
\includegraphics{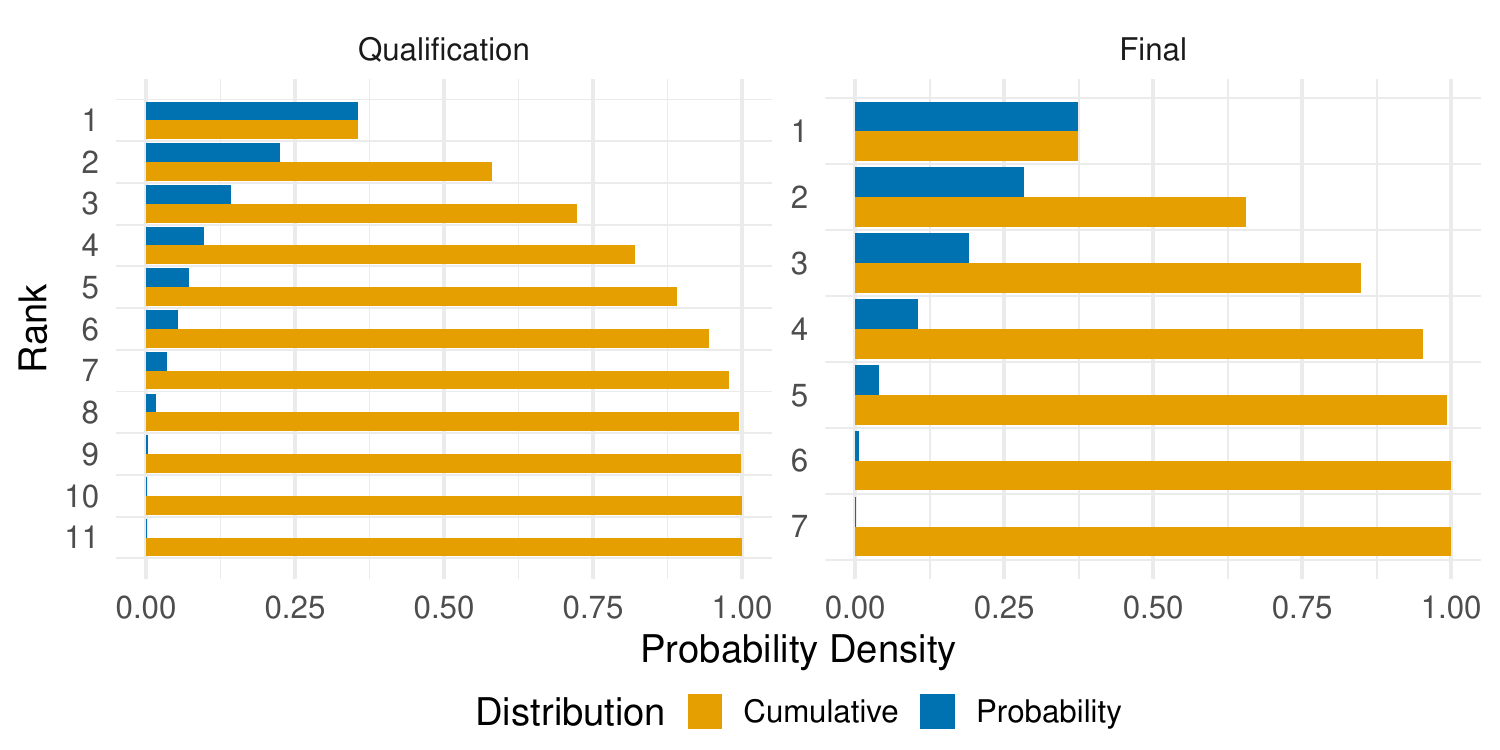}
\caption{\label{fig:fig2}The distributions for the probability of
finishing at each rank of both qualification and final rounds, given
that a climber wins any discipline, obtained from simulations. The
probability of finishing exactly at each given rank is coded blue,
whereas the probability of finishing at or better than each given rank
is coded orange. Results shown here are for Kendall rank correlation coefficients between bouldering and lead climbing of $0.526$ and $0.214$ for qualification and final rounds, with speed assumed independent of
both other events.}
\end{figure}

Our simulation results, as illustrated by Figure \ref{fig:fig2}, show
that a climber is almost guaranteed to finish in the top 8 of
qualification and advance to the final round if they win at least one of
the three climbing concentrations (99.5\% chance). Regarding the final round,
a climber is also very likely to claim a top 3 finish and bring home a
medal if they win any event (84.8\% chance). Moreover, we notice that if
a climber wins any discipline, they are also more likely to finish first
overall than any other positions in the eventual qualification and final
rankings. This shows how significant winning a discipline is to the
overall competition outcome for any given climber.

\begin{figure}
\centering
\includegraphics{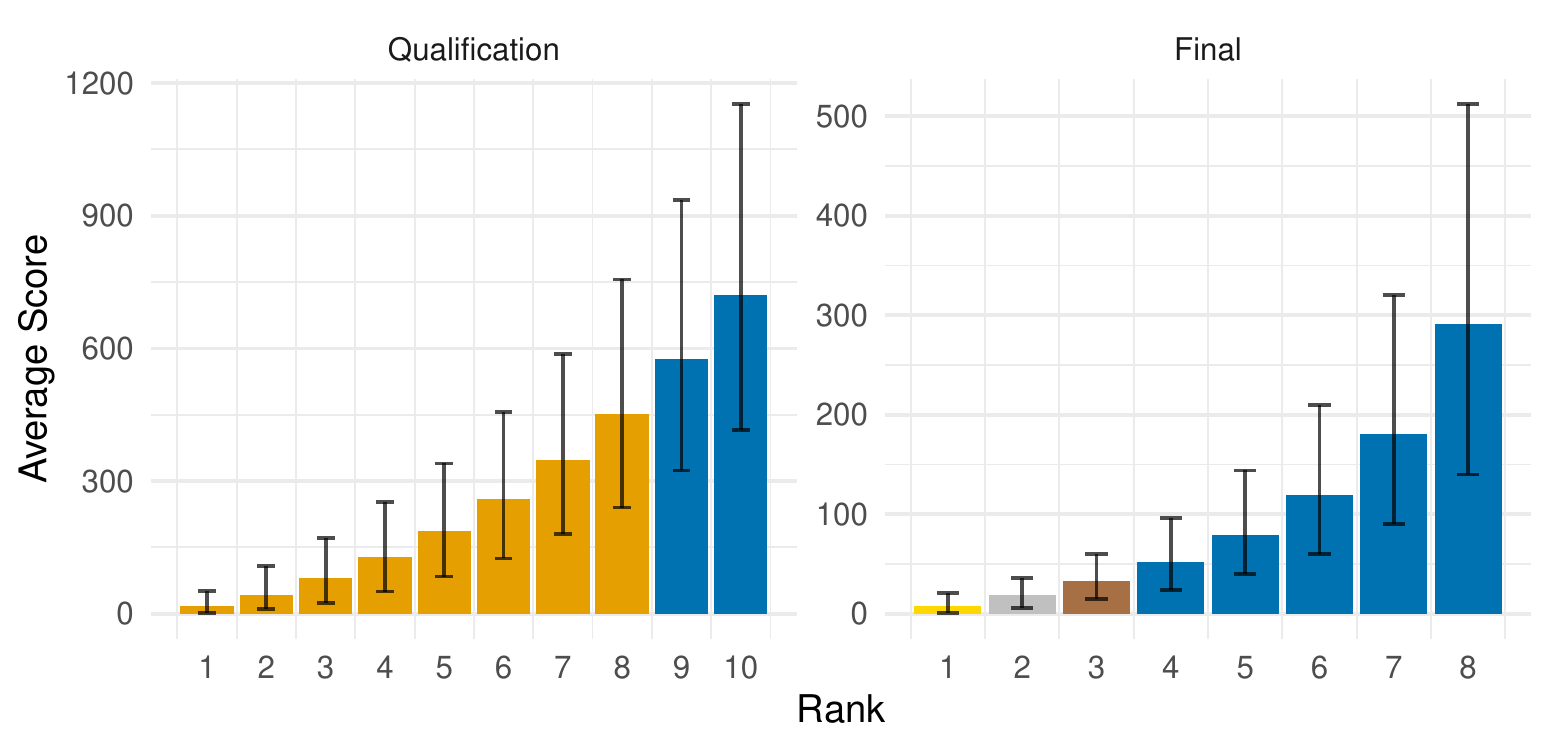}
\caption{\label{fig:fig3}The average scores (with 95\% confidence intervals) for the top 10 qualification ranks and all 8 final ranks, obtained from simulations. Results shown here are for Kendall rank correlation coefficients between bouldering and lead climbing of $0.526$ and $0.214$ for qualification and final rounds, with speed assumed independent of both other events.}
\end{figure}

In addition, we are interested in examining the distribution of the
total score for both qualification and final rounds. Figure
\ref{fig:fig3} is a summary of the expected score for each qualification
and final placement. According to our simulations, on average, the
qualification score that a contestant should aim for in order to move on
to the final round is 453 (for 8th rank). Furthermore, we observe that
in order to obtain a climbing medal, the average scores that put a
finalist in position to stand on the tri-level podium at the end of the
competition are 9, 19, and 33 for gold, silver, and bronze medals,
respectively.

\section{Data Analysis}
\label{sec:sec3}

\subsection{Correlations}
\label{combined-competition-format}

Throughout this section, we will be using data from the Women's Qualification at the 2020 Summer Olympics \citep{2021tokyo} as a case study for examining the relationship between the climber rankings in each individual discipline and the overall standings. The main attributes of this data are the name and nationality of the climbers; the finishing place of climbers in speed climbing, bouldering, and lead climbing; the total score (which equals the product of the discipline ranks); the overall placement; and the performance statistics associated with speed (race time), bouldering (tops-zones-attempts), and lead climbing (highest hold reached). We utilize this data to analyze the correlations between the event ranks and final table position, as well as to look at how often the final orderings change if one athlete is removed and the remaining climbers' ranks and total scores in each discipline are re-calculated. 

\begin{figure}
\centering
\includegraphics[width=0.9\linewidth]{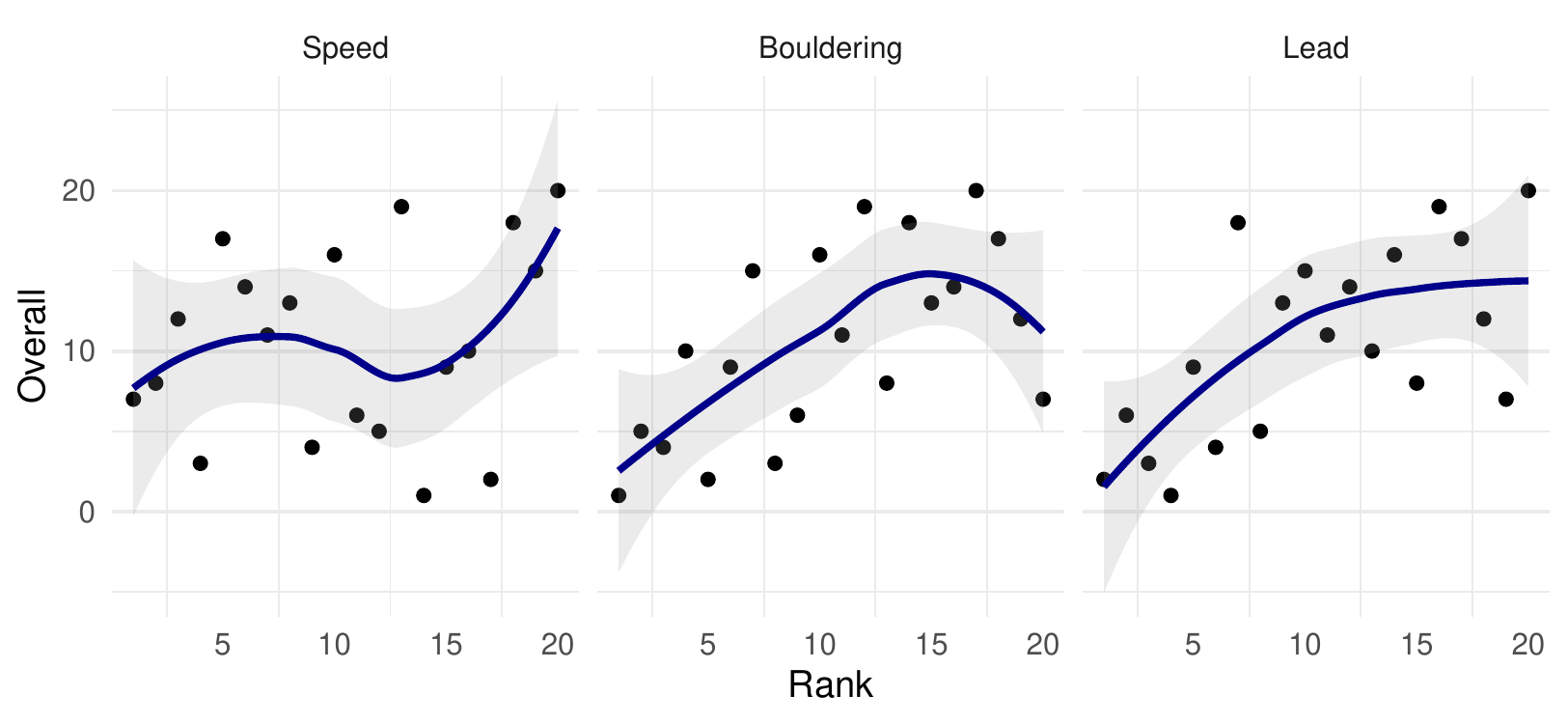}
\caption{\label{fig:fig4}Scatterplots with smoothed fitting curves and 95\% confidence intervals of overall rank and speed, bouldering, and lead ranks. Each data point represents a climber that competed in the Women’s Qualification at Tokyo 2020.}
\end{figure}

Figure \ref{fig:fig4} is a multi-panel matrix of scatterplots of the
ranks of the individual events and the final women's qualification
standings. We use Kendall's $\tau$ as our measure of
ordinal association between the ranked variables. Table \ref{tab:tab1}
shows the Kendall rank correlation coefficients between the overall rank
and the ranks of speed, bouldering, and lead disciplines; along with
corresponding \(95\%\) confidence intervals obtained from bootstrapping \citep{efron1986}.

\begin{table}[tbp]
  \caption{Kendall's $\tau$ values, along with correlation
test statistics, p-values, and bootstrapped 95\% confidence intervals,
for the overall rank and the rank of speed, bouldering, and lead
disciplines of Women's Qualification at Tokyo 2020.}%
  \label{tab:tab1}
  \centering
  \begin{tabular}[]{@{}lrrrr@{}}
\toprule
Discipline & Kendall's $\tau$ & Test Statistic & $p$-value & Bootstrapped
95\% CI \\
\midrule
Speed & $0.147$ & $109$ & $0.386$ & $(-0.191, 0.469)$ \\
Bouldering & $0.432$ & $136$ & $0.007$ & $(0.107, 0.691)$ \\
Lead & $0.463$ & $139$ & $0.004$ & $(0.112, 0.753)$ \\
\bottomrule
  \end{tabular}
\end{table}

It is evidently clear that there exists a fairly strong and positive
association between the final rank and the ranks of both bouldering
(\(\tau = 0.432\), \(p\)-value \(=0.007\), Bootstrapped \(95\%\) CI:
$(0.107, 0.691)$) and lead climbing (\(\tau = 0.463\), \(p\)-value
\(=0.004\), Bootstrapped \(95\%\) CI: $(0.112, 0.753)$). This implies
that climbers with high placements in both bouldering and lead also tend
to finish at a higher ranking spot overall.

Alternatively, the correlation with the final rank is not as strong for
speed climbing as the other two events (\(\tau = 0.147\)), and there is
insufficient evidence for an association between the rank of speed
climbing and the overall rank (\(p\)-value \(=0.386\), Bootstrapped
\(95\%\) CI: $(-0.191, 0.469)$). Thus, this offers evidence that speed
climbers are at a disadvantage under this three-discipline combined
format, compared to those with expertise in the other two
concentrations. Hence, it appears that the concerns of the climbers
mentioned in Section \ref{sec1pt1} may be valid.

In addition, we perform principal component analysis (PCA) to summarize
the correlations among a set of observed performance variables
associated with the three climbing disciplines. In particular, using the
data from the Women's Qualification at Tokyo 2020, we look at the
racing time (in seconds) for speed; the number of successfully completed
boulders (``tops''); and the number of holds reached for lead.

\begin{figure}

{\centering \includegraphics[width=0.8\linewidth]{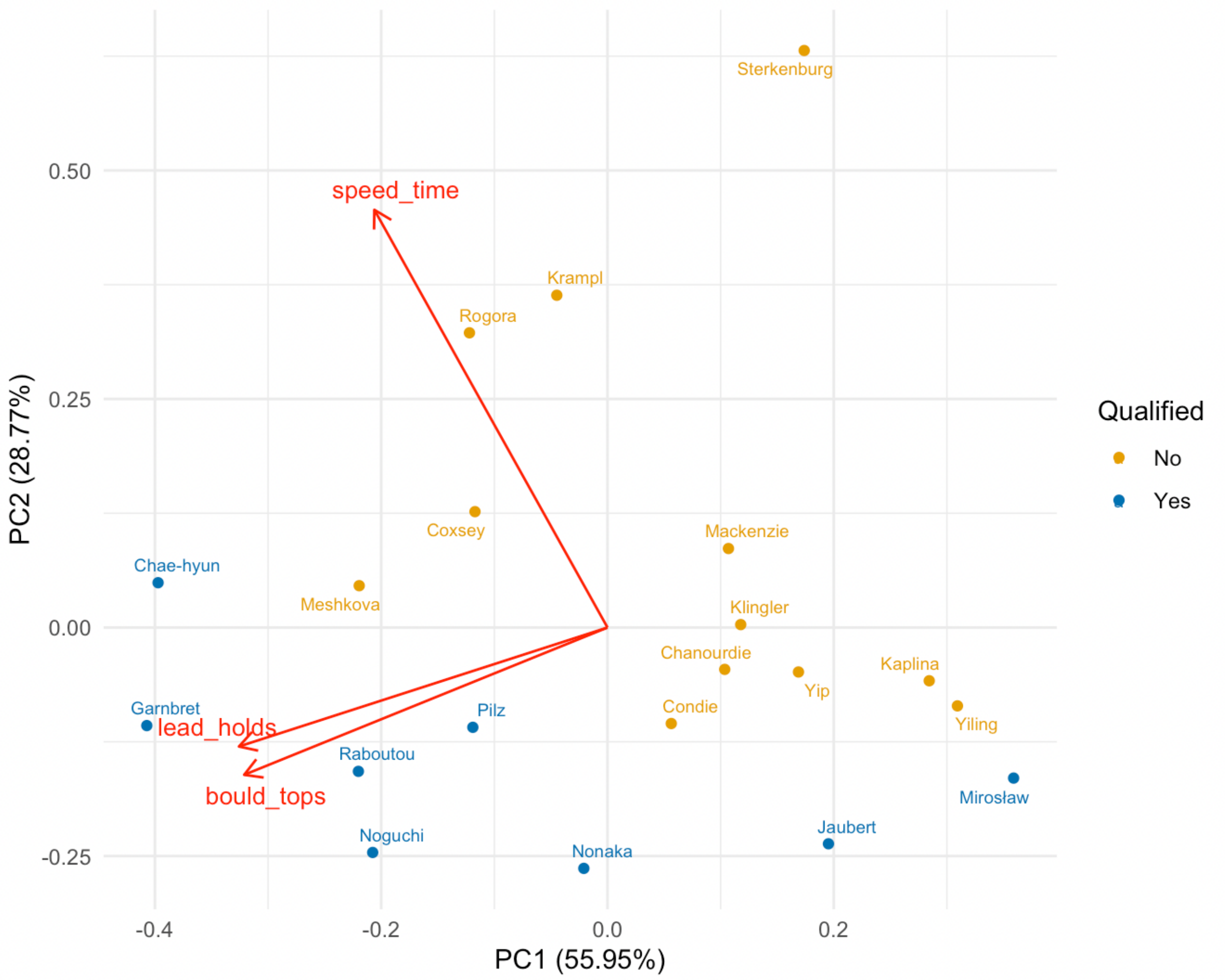} 

}

\caption{\label{fig:fig5}PCA Biplot for Women's Qualification at Tokyo 2020.}\label{fig:pca}
\end{figure}

Figure \ref{fig:fig5} is a PCA biplot showing the PC scores of the
climbers and loadings of the skill variables. We notice that the lead
and bouldering performances strongly influence PC1, while speed time is
the only variable contributing substantially to PC2, separated from the
other two skills. Moreover, since the two vectors representing the
loadings for the bouldering and lead performances are close and form a
small angle, the two variables they represent, bouldering tops and lead
holds reached, are positively correlated. This implies that a climber
that does well in bouldering is also very likely to deliver a good
performance in lead climbing.

\subsection{Leave-one-climber-out
Analysis}
\label{leave-one-climber-out-analysis}

Another question that we are interested in investigating is ``What would happen to the rankings if a single climber is removed?" There is a connection between this situation and the idea of independence of irrelevant alternatives (IIA) in social choice theory. The IIA criterion is a property of a voting system which states that after a winner is determined, if one of the losing candidates drops out and the votes are recounted, there should not be a change in the winner. First mentioned by \cite{arrow1951}, the IIA condition is also known as Luce's choice axiom \citep{luce1959} in probability theory, and it has had a number of applications in the fields of decision theory, economics, and psychology over the years. We notice a link between the concept of IIA and the topic of ranking system in sports. As an illustration, suppose we have 3 players A, B, and C participating in a competition. If A finishes in the first place and C is later disqualified and removed, A should still win. If the original winner (A) loses the modified competition (with C removed), then the Independence of Irrelevant Alternatives has been violated. For our particular case, this type of analysis can be helpful in examining the overall outcome for a climbing contest, specifically how a disqualification can affect the standings of medalists in the final round.

\begin{figure}
\centering
\includegraphics[width=0.95\linewidth]{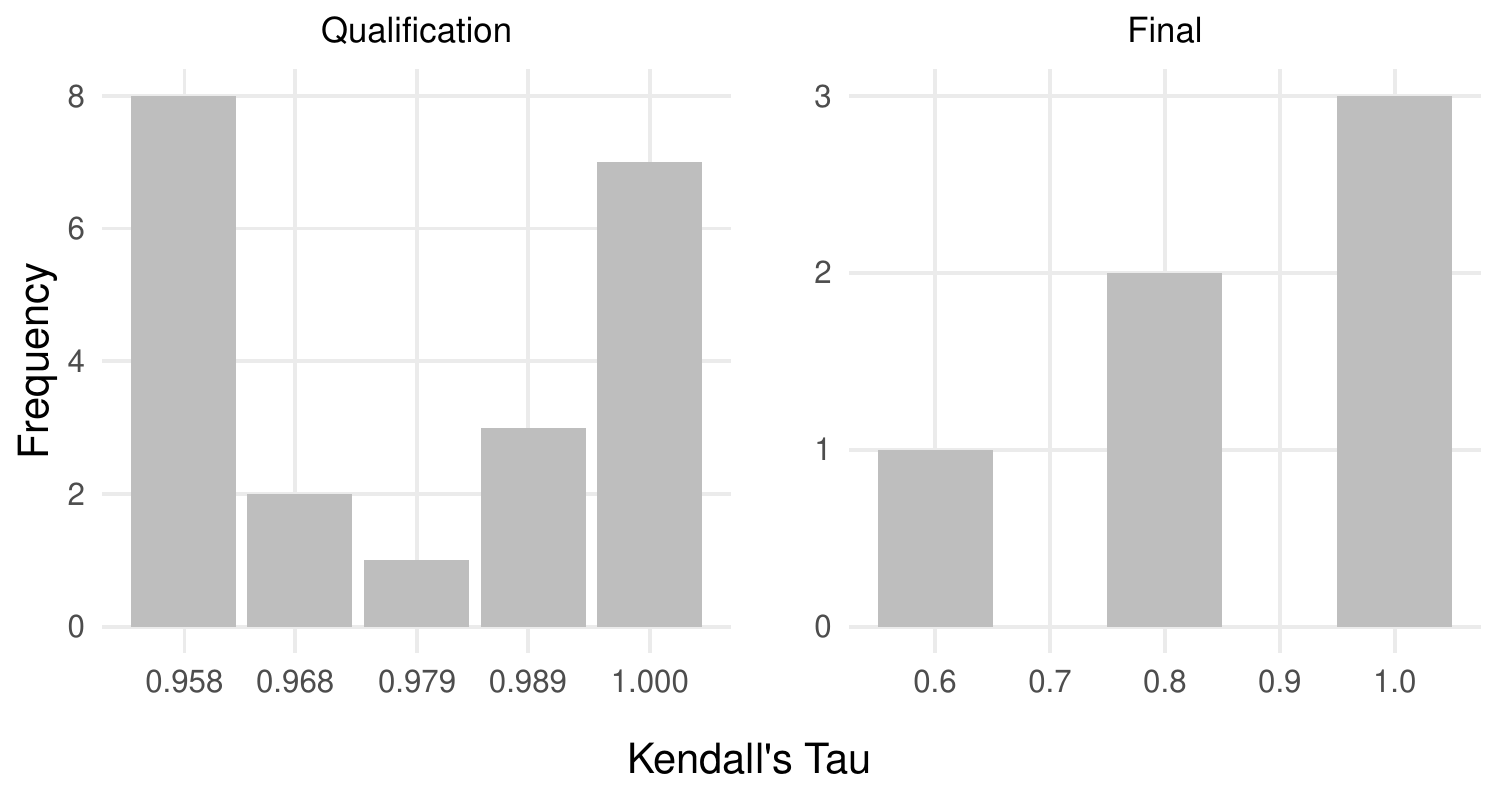}
\caption{\label{fig:fig6}This figure shows the distributions of Kendall's $\tau$ measuring the association between initial placements and new overall rankings after the removal of each rank for the women's qualification and final climbing rounds at the 2018 Youth Olympics. A perfect agreement of $\tau=1$ between the two sets of rankings is observed in 7 out of 21 rank elimination cases for qualification and 3 out of 6 for final.}
\end{figure}

To investigate whether the rank-product aggregation method of sport climbing violates social choice principles, we use data from the 2018 Youth Olympics women's climbing competition \citep{2018youth}. This event also implemented the combined format and rank-product scoring system, but consisted of 21 and 6 climbers competing in the qualification and final rounds, respectively, rather than 20 and 8 like the Tokyo 2020 Olympics. The analysis is performed as follows. After an athlete is dropped (by their original placement), the ranks for each discipline of the remaining players are re-calculated. The new total scores are then obtained as before, by multiplying the three event ranks, which then determines the new overall finishing positions. For each rank elimination case, the association between the original and modified standings is summarized, and a distribution of Kendall correlation between the two sets of rankings is obtained for each round. Figure \ref{fig:fig6} reveals that there is not always a perfect concordance between the original orderings and new overall placements of the remaining climbers after the person with a specified rank is removed. In particular, a non-perfect agreement between the two sets of rankings is observed in two-thirds and one-half of the rank exclusion instances for the qualification and final rounds, respectively.

\begin{figure}
\centering
\includegraphics[width=0.95\linewidth]{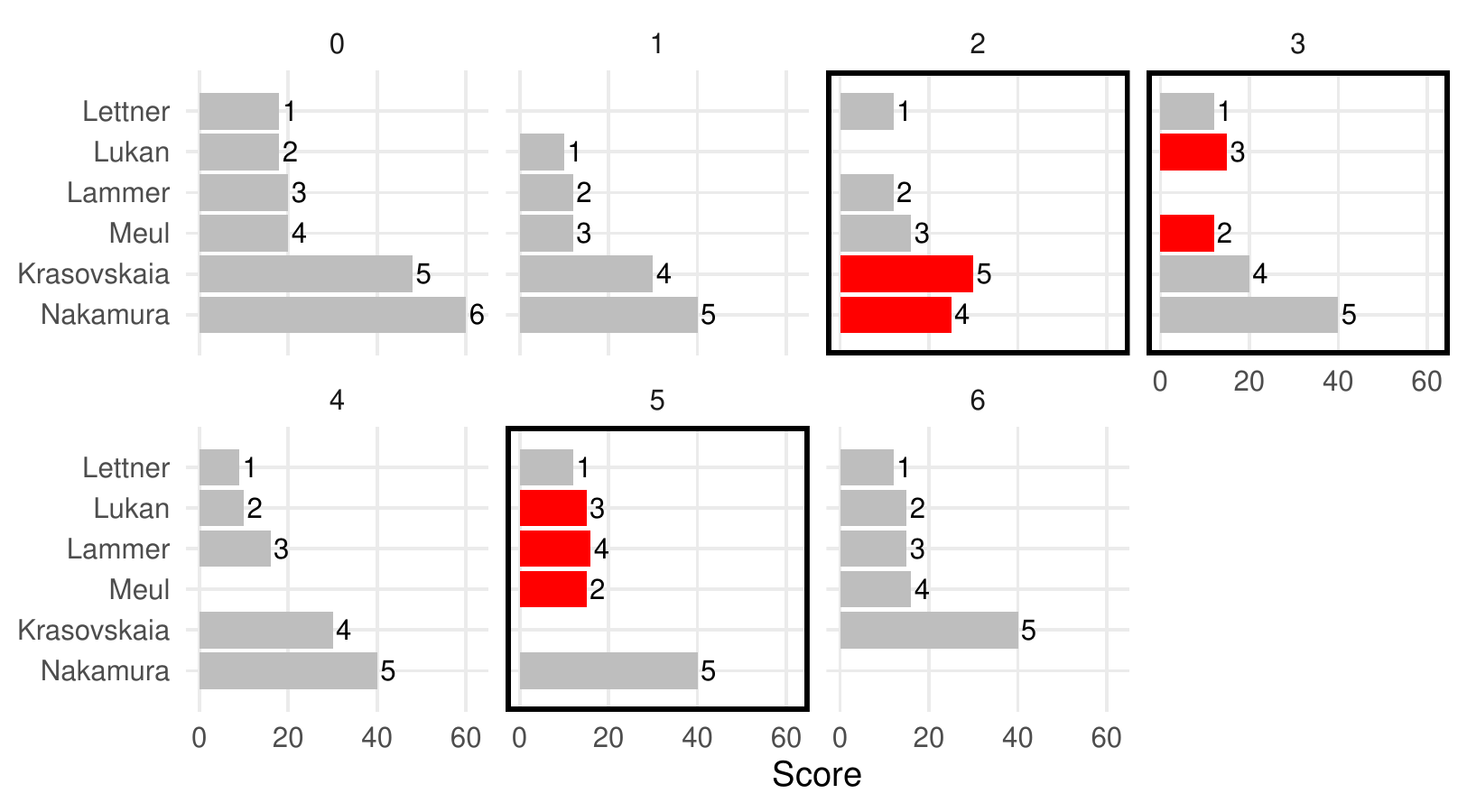}
\caption{\label{fig:fig7}This figure illustrates the changes to the 2018
Youth Olympics Women's Final rankings when each climber is left out.
Each panel represents the rank of the drop-out athlete, with 0 being the
original final results. Each case with a change in rank orderings is
highlighted by a black panel border, and any player with a rank change
is represented by a red-filled bar.}
\end{figure}

Furthermore, Figure \ref{fig:fig7} shows the modified versions of the rankings after
each ranked climber is excluded for the Women's Final at the 2018 Youth
Olympics. We observe from this plot that removing a single contestant
changes the rankings considerably, especially in terms of the order of
medalists. One particular interesting case is where an athlete's
finishing position changes when someone who originally finished behind them drops
out. This situation is illustrated by panel 5 of the women's
competition, where the fifth-place climber, Krasovskaia, was excluded;
and Meul, whose actual placement was fourth, moved up to the second
spot and would have claimed the silver medal. Moreover, this clearly
demonstrates that the rank-product scoring method of Olympics sport
climbing violates the independence of irrelevant alternatives criterion. 
To reiterate, Meul actually finished fourth in the 2018 Youth Olympics. 
However, \textbf{with the exact same performance} in the final round, 
she would have finished in second place, winning a medal no less, if Krasovskaia, 
who finished behind Meul in fifth, had simply not competed.



\section{Conclusion and Discussion}
\label{sec:sec4}

In this paper, we examined the general features of the rank-product
scoring system of sport climbing, in particular, the advancement
probabilities and scores for the climbers. Most importantly, through
analyses of historical climbing data, we pointed out several problems of
the combined competition format and rank-product scoring system of sport
climbing. First, combining the three disciplines speed, bouldering, and
lead together into a ``triathlon'' format is putting speed
climbers in an unfavorable position. Second, the sport climbing rank-product
aggregation method violates the independence of irrelevant alternatives. 
As such, there is a dependency on irrelevant parties in this scoring scheme, as the
orderings of medalists can be affected with a drop-out of a lower-ranked
climber.

While we did not attempt to propose a new rank aggregation method to replace
rank-product scoring in combined climbing contest, we instead have a
suggestion to modify the overall structure of the competition. In
particular, we suggest that speed climbing should have its own set of
medals in future tournaments, whereas bouldering and lead
climbing can be amalgamated into one event. In fact, it was confirmed that
in the next Summer Olympics held in Paris in 2024, there will be two
separate contests and two sets of climbing medals for each gender event:
combined lead and bouldering, and speed-only \citep{goh2020}. This is
consistent with what we have shown, as bouldering and lead climbing
performances are highly correlated with each other and with the overall
result, whereas speed climbing should be separated from the combined
competition format.

\vspace{-4mm}

\section*{Acknowledgements}
\label{acknowledgements}

The authors would like to acknowledge the organizers of the 2021 UConn
Sports Analytics Symposium and the 2021 Carnegie Mellon Sports Analytics
Conference (CMSAC 2021) for the opportunities to present this work at
its early stage and receive feedback. In addition, we would like to
express our special thanks to the anonymous reviewers of the
Reproducible Research Competition at CMSAC 2021 for the insightful
comments and suggestions that led to significant improvements in the
paper.

\section*{Supplementary Material}
\label{supplementary-material}

All data and code for reproducing the analyses presented in this
manuscript are publicly available on GitHub at
\url{https://github.com/qntkhvn/climbing}.

\bibliographystyle{jds}
\bibliography{climbing_jds}

\end{document}